 \documentclass[12pt,preprint]{aastex}

\usepackage{graphicx}

\def \apj {ApJ}

\def \apjl {ApJL}

\def \aap {A\&A}

\newcommand{\citeN}[1]{\citeauthor{#1} (\citeyear{#1})}
\newcommand{\citeNP}[1]{\citeauthor{#1} \citeyear{#1}}

\shortauthors{Socas-Navarro}
\shorttitle{Are Electric Currents Heating the Chromosphere?}

\begin{document}

\title{Are Electric Currents Heating the Magnetic Chromosphere?}

\author{H. Socas-Navarro}
   	\affil{High Altitude Observatory, NCAR\thanks{The National Center
	for Atmospheric Research (NCAR) is sponsored by the National Science
	Foundation.}, 3450 Mitchell Lane, Boulder, CO 80307-3000, USA}
	\email{navarro@ucar.edu}

\date{}%

\begin{abstract}
This paper presents an analysis of three-dimensional vector currents and
temperatures observed in a sunspot from the photosphere to the chromosphere,
spanning a range of heights of approximately 1500~km. With this unique
dataset, based on novel spectro-polarimetric observations of the 850~nm
spectral region, it is possible to conduct for the first time an empirical
study of the relation between currents and chromospheric heating. It is shown
that, while resistive current dissipation contributes to heat the sunspot
chromosphere, it is not the dominant factor. The heating effect of current
dissipation is more important in the penumbra of the sunspot, but even there
it is still a relatively modest contribution. 
\end{abstract}
   


Currents are thought to play an important role for the energetics of the
chromosphere and the corona of the Sun and, by extrapolation, of cool stars
in general.  Essentially, three types of processes have been proposed to
explain the chromospheric heating problem (\citeNP{UPR91}): a)heating by
upward-propagating waves (acoustic and/or magnetic, \citeNP{CS92};
\citeNP{A47}); b)energy release by magnetic reconnection(\citeNP{P88}) and
c)resistive dissipation of electric
currents (\citeNP{RM84}; \citeNP{G04}). Unfortunately, the remote detection of
solar electric currents is very challenging and requires accurate 3D vector
magnetometry to solve, at each spatial point, the vector equation $\vec
j=\nabla \times \vec B$. Because of this, the measurement of currents in the
chromosphere has been essentially limited to the recent work of
\citeN{SLW+03}, based on the \ion{He}{1}
multiplet at 1083~nm. These lines have the advantage that they can be
analyzed with a simplified modeling technique and one does not need to deal
with the complications of detailed non-LTE radiative transfer
calculations. The drawback is that this simplified analysis does not allow
the authors to obtain temperatures or even to determine empirically the
height of measurement (which varies from pixel to pixel). While that work was
an important step forward, it could not establish whether electric currents
constitute an effective heating source.


The observations used in this work were acquired on June 2004 with the new
instrument SPINOR (Spectro-Polarimeter for Infrared and Optical Regions,
\citeNP{SNEP+05a}). The dataset, consisting of a spectro-polarimetric scan of
NOAA~0634 in the 849.8 and 854.2~nm spectral region, was analyzed with the
non-LTE inversion code of \citeN{SNTBRC01}. This produced a 3D reconstruction
of the temperature, velocity and magnetic field vector from the photosphere
to the chromosphere (a more detailed account is given in \citeNP{SN05c}).

Having the 3D structure of the vector magnetic field, one can
compute the vector current density as the curl of the field.
Figure 1 shows maps of the current density at different heights in
the atmosphere. The strongest currents are organized in filamentary
structures, at least when viewed in these 2D representations (horizontal
cuts). In 3D their appearance is more reminiscent of "sheets". The western
boundary of the sunspot (right of the images) exhibits a very prominent
system of strong currents, especially in the photosphere. As I discuss
below, these currents are not aligned with the vector field, which implies
that the field is "forced" (i.e., non-force-free). The magnetic topology
in the area surrounding the spot is very different on both sides of it, as
evidenced by larger-field magnetograms (not shown in the figure). While the
east side is almost devoid of strong flux, the west side harbors a very
intricate pattern of strong magnetic fields. It is interesting to note that
H$\alpha$ images of this active region show considerable activity on the west
side very near the sunspot. Moreover, we identified two Ellerman
bombs (\citeNP{SNEP+05b}) located off of the edge of the western penumbra. The
combination of 
H$\alpha$ emission, Ellerman bombs and complex field topology is a symptom of
ongoing magnetic reconnection, which is consistent with the presence of
strong currents such as those shown in the figure. 

Figure 2 shows the angle between the currents and the magnetic field
lines. In a force-free field this angle would be either 0 (parallel) or 180
(anti-parallel) degrees everywhere. We can see that, as one would expect, the
field is more force-free in the chromosphere than in the
photosphere. However, even the higher chromosphere is far from being entirely
force-free. Only the inner part of the eastern penumbra is fairly force-free
and exhibits a pattern of alternating parallel/anti-parallel field-aligned
currents running along the penumbral filaments. The filamentary structure
extends outwards, even beyond the limit of the visible penumbra, but with
significant inclinations with respect to the field lines. The same pattern of
incoming/outgoing non-field-aligned currents is visible in the North-East
(upper-left) quadrant of the photospheric penumbra. 

The reliability of the results presented in Figs~1 and~2 is supported by
several noteworthy arguments: The inversion code has been extensively tested
(\citeNP{SNTBRC01}) and found to produce a good representation of the average
thermal and magnetic conditions in the spatial pixel. The parameters
reconstructed by the tomography (temperature, magnetic field, currents, etc)
exhibit smooth pixel-to-pixel variations and spatial coherence, even though
each pixel was inverted independently. Particularly interesting is the fact
that currents appear as filaments in the maps (or sheets in 3D), which would
be difficult to produce by artifacts of the analysis. Furthermore, the
results obtained are physically sensible (the field is more force-free in the
chromosphere than in the photosphere, the connection of strong currents with
H$\alpha$ emission and Ellerman bombs, the filamentary pattern extending even
beyond the visible penumbra, etc). But perhaps the most convincing test for
the accuracy of the tomography is that of the divergence of the field which,
according to the Maxwell equations, should be zero everywhere. I have
verified that this divergence, computed using boxes of 1.4 Mm side, is always
small when compared to $B/l$ (the ratio of the field strength over the length
of the box), with an average absolute value of 1.8\% of that ratio. Again,
this would be very difficult to produce by random or systematic analysis
errors.

The most important source of error in obtaining the currents is the
$z$-scale, i.e. the geometrical height of each point in the 3D grid. We
carried out a very thorough treatment of the height scale, including a
correction of the Wilson depression by matching the total (gas and magnetic)
pressure at neighboring points (\citeNP{SA05}). Still, uncertainties are to be
expected of up to 50-100~km through the photosphere and 200-300~km in the
chromosphere. A Montecarlo-type error analysis was conducted by adding random
perturbations to the $z$-scale at each ($x$,$y$)-point and recomputing the
vector current density. The maps obtained did not change significantly by
perturbations at this level (with relative errors of $\sim$10\%), which is a
natural consequence of the field gradients being relatively smooth in the
sunspot.  

It is also important to consider the effects of unresolved structure in the
observed pixel, since penumbral filaments are known to be organized on small
scales. In such situations the inversion procedure retrieves an average of
the thermal and magnetic conditions in the resolution element. By taking
volume integrals on both sides of the equation above, and using the
mathematical property that the integral of the curl is the curl of the
integral, it is straightforward to see that one then obtains the average
current in the resolution element.

Let us now turn to the thermal structure of the sunspot and its relation to
electric currents. Figure~3 shows horizontal cuts of the electron temperature
at several heights. We can see that the chromosphere exhibits "hot patches",
with typical size scales of $\sim$5~Mm, embedded in a cooler medium. When
this figure is compared to the current density maps discussed above, we can
see that they are fundamentally different and there is no obvious connection
between chromospheric hot spots and the strong current sheets. A more
quantitative analysis is presented in Fig~4, which shows the spatial
correlation between currents and temperatures. Notice that, even though the
overall correlations are very weak, they are nearly always positive. This is
an indication that current dissipation is indeed contributing to the
chromospheric heating above sunspots, but it is not the dominant
mechanism. The correlation coefficient increases by about a factor of two
(but still remains weak) if we restrict the analysis to the penumbral region
(right panels), suggesting that the effect of electric currents is more
important in the penumbra. It has been proposed (see e.g., \citeNP{G04}) that
only the component perpendicular to the field (the so-called Pedersen
currents) is relevant for the heating of the upper atmosphere. The
correlation curves for the entire region (left panels in the figure) do not
generally show a clear distinction between both components. However, the
penumbra (right panels) exhibits a slightly better correlation with the
Pedersen component, in agreement with that statement. 

The results presented in this letter do not just argue against resistive
current dissipation as a dominant source for chromospheric heating. Since
magnetic reconnection is associated with strong current sheets, we can also
conclude that reconnection must not be important, at least in the relatively
simple magnetic topology of sunspots (the situation is likely to be different
in more complex regions, such as the emerging flux scenario analyzed by
\citeNP{SLW+03}). This leaves only one of the three types of mechanisms
mentioned above, namely the dissipation of (acoustic or magnetic) waves. In a
recent work, \citeN{FC05} claim that acoustic waves do not 
carry enough energy to sustain the heating rate of the upper solar
atmosphere. The combination of their results with the ones presented here
appears to point directly to Alfv\' en waves as the dominant heating source,
at least in sunspots.



\clearpage

\begin{thebibliography}{11}
\expandafter\ifx\csname natexlab\endcsname\relax\def\natexlab#1{#1}\fi

\bibitem[{{Alfv{\' e}n}(1947)}]{A47}
{Alfv{\' e}n}, H. 1947, \mnras, 107, 211

\bibitem[{{Carlsson} \& {Stein}(1992)}]{CS92}
{Carlsson}, M., \& {Stein}, R.~F. 1992, \apjl, 397, L59

\bibitem[{{Fossum} \& {Carlsson} (2005)}]{FC05}
{Fossum}, , \& {Carlsson}, M. 2005, \nat, 435, 919

\bibitem[{{Goodman}(2004)}]{G04}
{Goodman}, M.~L. 2004, \aap, 424, 691

\bibitem[{{Parker}(1988)}]{P88}
{Parker}, E.~N. 1988, \apj, 330, 474

\bibitem[{{Rabin} \& {Moore}(1984)}]{RM84}
{Rabin}, D., \& {Moore}, R. 1984, \apj, 285, 359

\bibitem[{{S{\' a}nchez Almeida}(2005)}]{SA05}
{S{\' a}nchez Almeida}, J. 2005, \apj, 622, 1292

\bibitem[{{Socas-Navarro}(2005)}]{SN05c}
{Socas-Navarro}, H. 2005, \apjl, {\it submitted}

\bibitem[{{Socas-Navarro} {et~al.}(2005{\natexlab{a}}){Socas-Navarro},
  {Elmore}, {Pietarila}, {Darnell}, {Tomczyk}, \& {Lites}}]{SNEP+05a}
{Socas-Navarro}, H., {Elmore}, D., {Pietarila}, A., {Darnell}, T., {Tomczyk},
  S., \& {Lites}, B. 2005{\natexlab{a}}, Solar Physics, {\it submitted}

\bibitem[{{Socas-Navarro} {et~al.}(2005{\natexlab{b}}){Socas-Navarro},
  {Elmore}, {Pietarila}, {Lites}, {Manso Sainz}, \& {Mart\'\i nez
  Pillet}}]{SNEP+05b}
{Socas-Navarro}, H., {Elmore}, D., {Pietarila}, A., {Lites}, B., {Manso Sainz},
  R., \& {Mart\'\i nez Pillet}, V. 2005{\natexlab{b}}, Solar Physics, {\it in
  preparation}

\bibitem[{{Socas-Navarro} {et~al.}(2001){Socas-Navarro}, {Trujillo Bueno}, \&
  {Ruiz Cobo}}]{SNTBRC01}
{Socas-Navarro}, H., {Trujillo Bueno}, J., \& {Ruiz Cobo}, B. 2001, \apj, 550,
  1102

\bibitem[{{Solanki} {et~al.}(2003){Solanki}, {Lagg}, {Woch}, {Krupp}, \&
  {Collados}}]{SLW+03}
{Solanki}, S.~K., {Lagg}, A., {Woch}, J., {Krupp}, N., \& {Collados}, M. 2003,
  \nat, 425, 692

\bibitem[{{Ulmschneider} {et~al.}(1991){Ulmschneider}, {Priest}, \&
  {Rosner}}]{UPR91}
{Ulmschneider}, P., {Priest}, E.~R., \& {Rosner}, R., eds. 1991, {Mechanisms of
  Chromospheric and Coronal Heating}

\end{thebibliography}

\clearpage

\begin{figure*}
\plotone{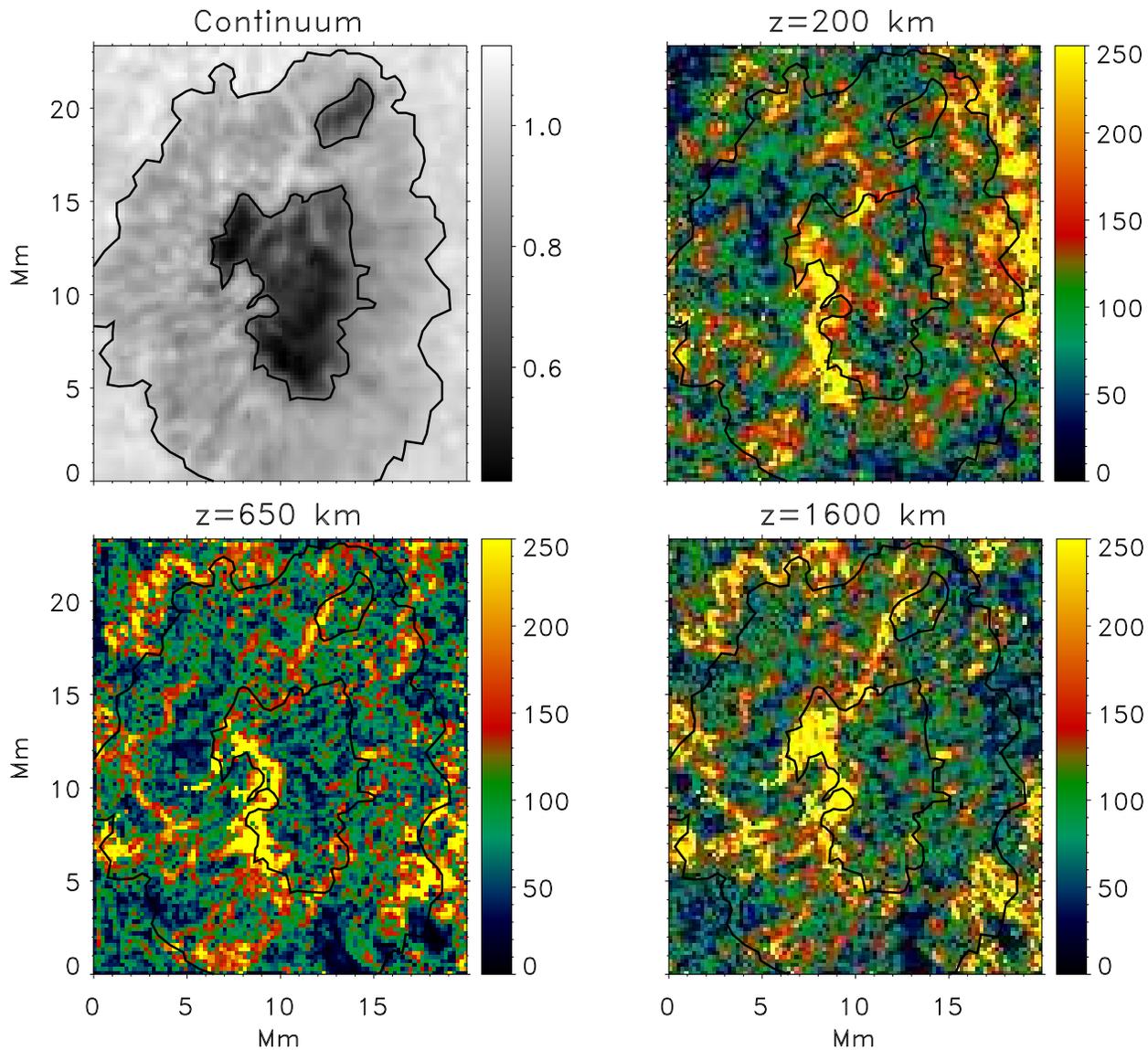}
\caption{Upper left: Continuum image of the sunspot analyzed in this work 
(in units of the average quiet Sun). North is up and East is left of the 
image. The rest of the panels present current density maps (units are 10$^3$
A/km$^2$) at $z$=200, 650 and 1600~km. The $z$=0 level is the height at which
the continuum optical depth at 500~nm reaches unity ($\tau_{500}$=1) in the
quiet Sun.   
\label{fig:curr}
}
\end{figure*}

\clearpage

\begin{figure*}
\plotone{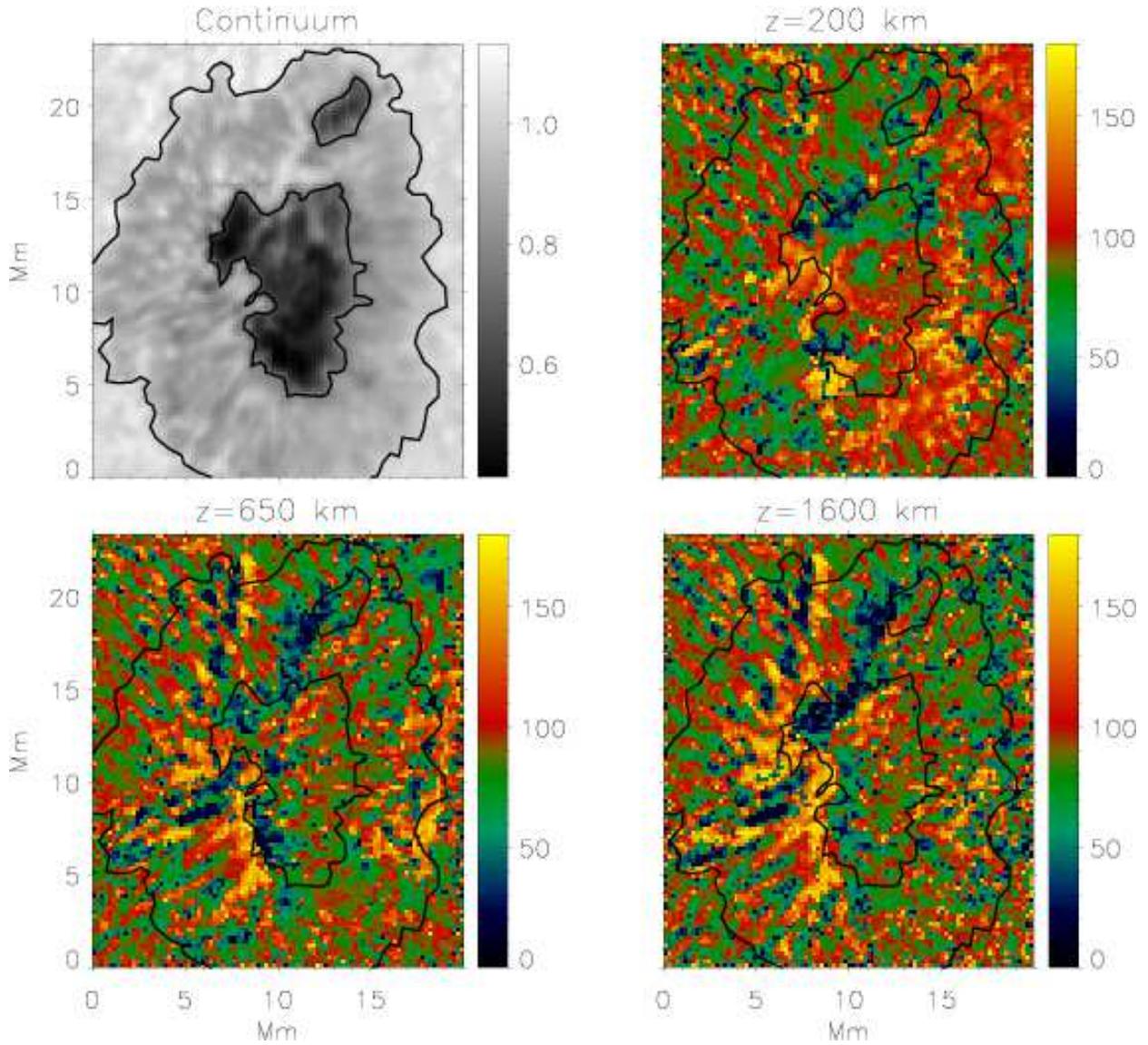}
\caption{Upper left: Continuum image of the sunspot analyzed in this work (in
  units of the average quiet Sun). North is up and East is left of the
  image. The rest of the panels present the angles (degrees) between the
  vector current and the magnetic field.  
\label{fig:angles}
}
\end{figure*}

\clearpage

\begin{figure*}
\plotone{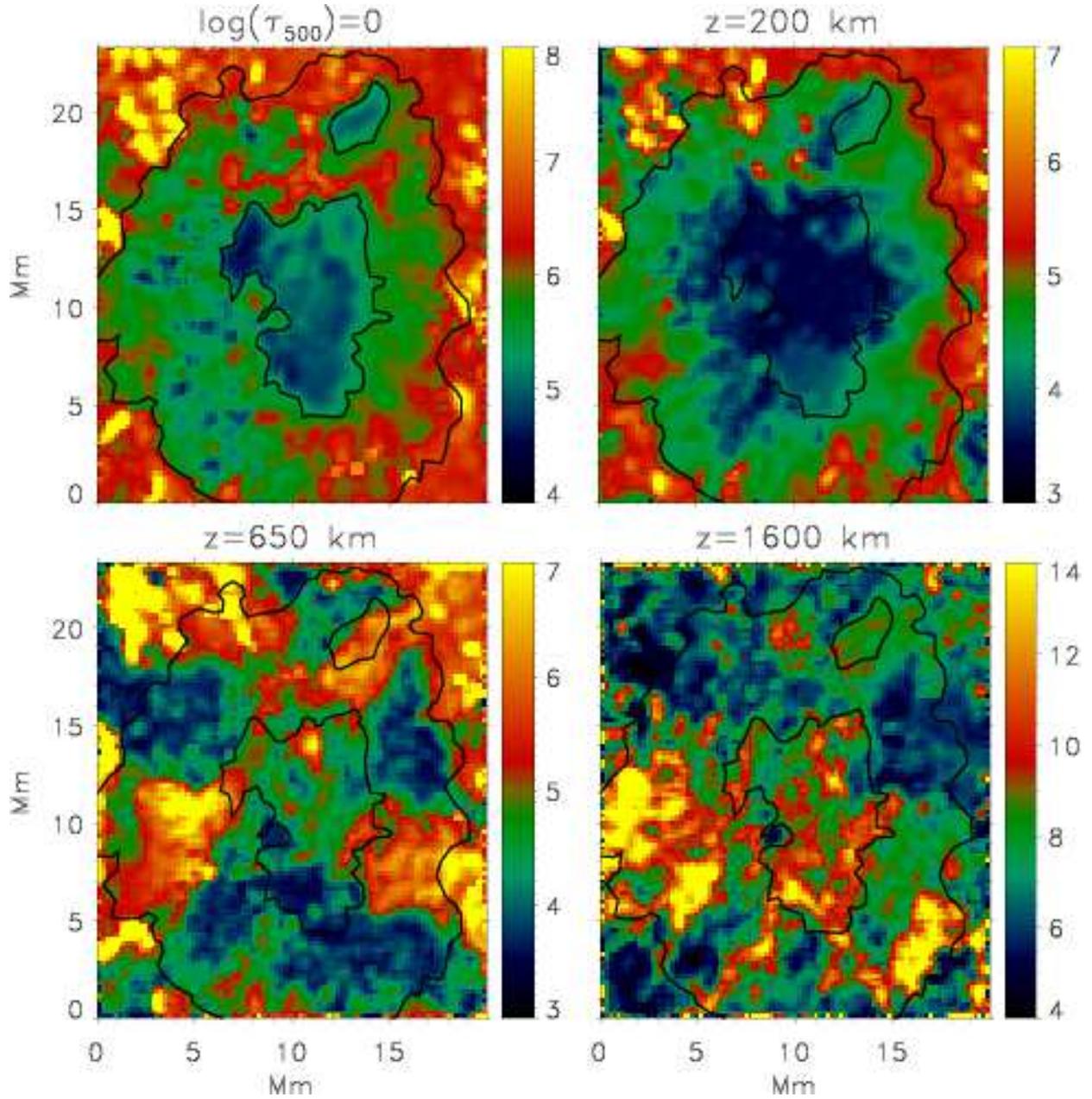}
\caption{Temperature maps (in kK) at different heights in the
  atmosphere. Upper left: Temperatures at optical depth $\tau_{500}$=1. Rest
  of the panels: Temperatures at geometrical heights $z$=200, 650 and
  1600~km, respectively.
\label{fig:temps}
}
\end{figure*}

\clearpage

\begin{figure*}
\plotone{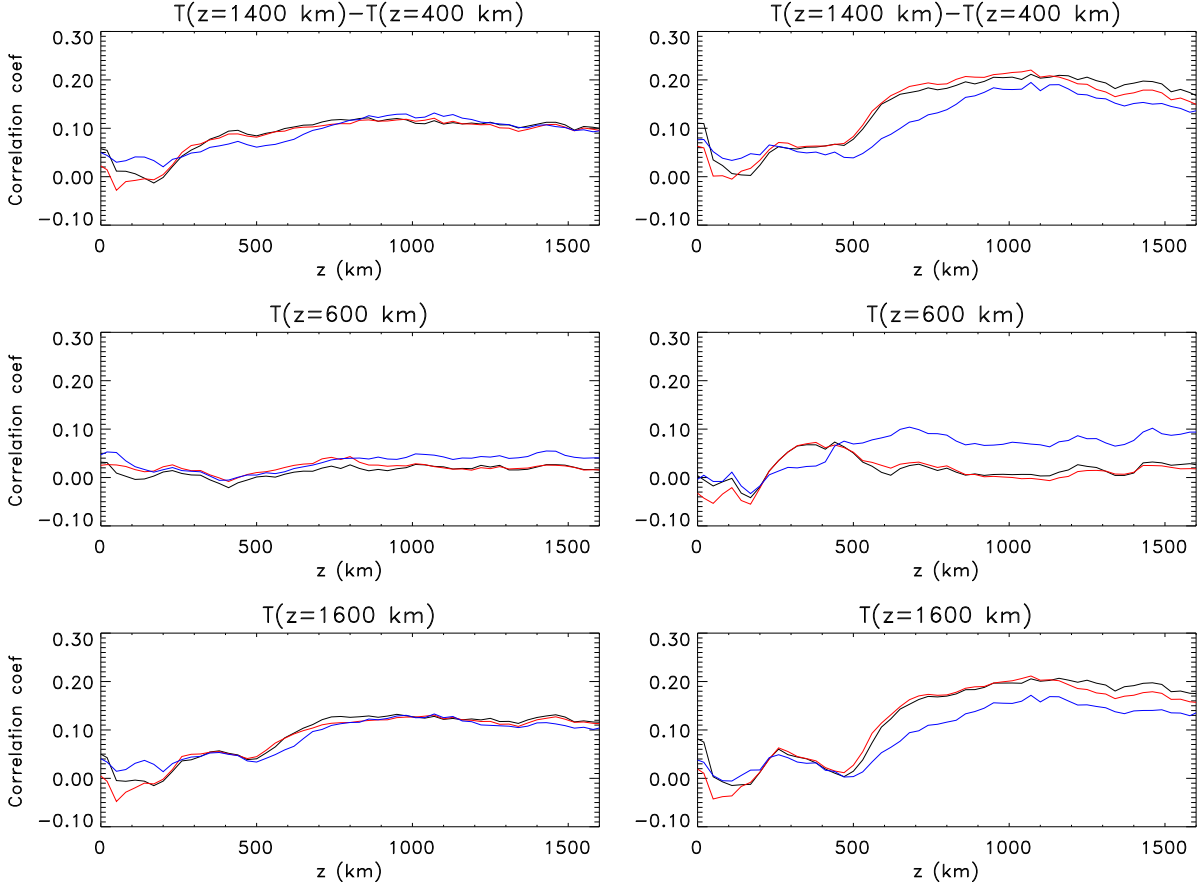}
\caption{Pearson's correlation coefficient between the spatial distribution
  of temperatures at the geometrical height indicated by the label of each
  plot and the current densities at height $z$. The black line represents the
  modulus of the vector current density, whereas the red and blue lines
  represent the components perpendicular and parallel to the field,
  respectively. The upper panels refer to the temperature gradient between
  the upper photosphere and the chromosphere. Under strict radiative
  equilibrium conditions, this gradient should be slightly negative and the
  temperature would decrease monotonically with height. Therefore, the
  patches with positive temperature gradient are regions with important
  heating activity. The middle and lower panels depict the correlations
  between currents and temperatures at the lower and upper chromosphere,
  respectively. The left panels include the entire dataset, whereas the right
  panels are restricted to the sunspot penumbra only.  
\label{fig:correls}
}
\end{figure*}

\end{document}